\documentclass[prl,aps,twocolumn,superscriptaddress]{revtex4}
\usepackage{epsfig}
\usepackage{import}
\usepackage{amsmath}
\input{epsf}
\usepackage{psfrag}
\usepackage{xcolor}
\usepackage{slashed}
\usepackage{hyperref}
\usepackage{ulem}
\usepackage{CJKutf8}

\begin{document}


\title{Probing quark orbital angular momentum at EIC and EicC}
 
\author{Shohini Bhattacharya}
\affiliation{RIKEN BNL Research Center, Brookhaven National Laboratory, Upton, NY 11973, USA}

\author{Duxin Zheng}
\affiliation{Shandong Institute of Advanced Technology, Jinan, Shandong, 250100, China}

\author{Jian Zhou}
\affiliation{School of Physics and Key Laboratory of Particle Physics and Particle Irradiation (MOE), Shandong University, QingDao, Shandong, 266237, China}

\begin{abstract}
We propose to detect signals from quark orbital angular momentum (OAM) through exclusive $\pi^0$ production in electron-(longitudinally-polarized) proton collisions. Our analysis demonstrates that the $\sin 2\phi$ azimuthal angular correlation between the transverse momentum of the scattered electron and the recoil proton serves as a sensitive probe of quark OAM. Additionally, we present a numerical estimate of the asymmetry associated with this correlation for the kinematics accessible at EIC and EicC. This study aims to pave the way for the first experimental study of quark OAM in relation to the Jaffe-Manohar spin sum rule.

\end{abstract}

\maketitle

\date{\today}

{\it  1.  Introduction}---The exploration of nucleon spin structure, sparked by the revelation of the ``spin crisis", has developed into a captivating research field over the past three decades. A central goal of this field is to comprehend the nucleon's spin in terms of contributions from its underlying partons. Significant progress has been made in deciphering this partonic content of nucleon spin, particularly in constraining contributions from quark and gluon spins in the moderate and large $x$ regions through measurements of parton helicity distributions at accelerator facilities worldwide~\cite{STAR:2014wox,deFlorian:2014yva,Nocera:2014gqa,STAR:2021mqa}. The upcoming Electron-Ion Collider in the US and China (EIC and EicC)~\cite{AbdulKhalek:2021gbh, Anderle:2021wcy} is expected to play a crucial role in precisely determining the gluon helicity distribution at small $x$. While parton helicities represent a significant fraction of nucleon spin, there remains ample opportunity to investigate the contribution of parton Orbital Angular Momentum (OAM) to nucleon spin, constituting another key objective of the EIC and EicC.
\begin{figure}
    \centering
\includegraphics[scale=0.15]{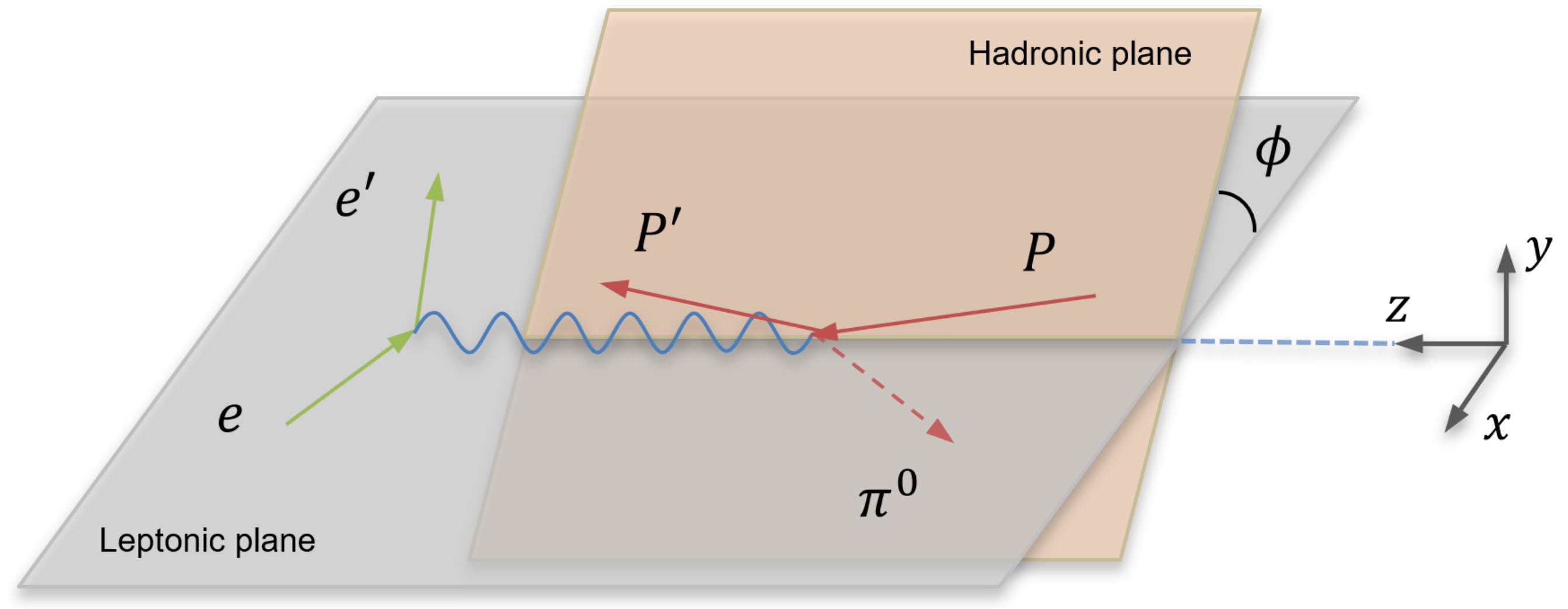}
   \caption{An illustration of  exclusive $\pi^0$ production.}
\label{illus}
\end{figure}

In an interacting theory like Quantum Chromodynamics (QCD), two types of OAMs exist: the kinetic type (Ji's type) and the canonical type (Jaffe-Manohar's type).  The difference between the two definitions of OAM in QCD can be attributed to the gauge potential term. In practice, the kinetic OAM of quarks and gluons is determined by subtracting their helicity contributions from the total angular momentum contributions, which can be accessed through hard exclusive processes~\cite{Ji:1996ek,Ji:1996nm}. However, extracting Jaffe-Manohar type parton OAM~\cite{Jaffe:1989jz}, or equivalently canonical OAM, in high-energy scattering processes poses a significant experimental challenge. Progress in this direction was limited until a connection between parton OAM and Wigner distribution functions~\cite{Belitsky:2003nz}, or equivalently, Generalized Transverse Momentum-dependent Distributions (GTMDs)~\cite{Meissner:2009ww}, was revealed about a decade ago. For the quark case, this connection is given by~\cite{Lorce:2011kd,Hatta:2011ku,Lorce:2011ni},
\begin{eqnarray} 
L^{q}(x,\xi)=-\int d^2 k_\perp \frac{ 
 k_\perp^2}{M^2} F_{1,4}^{q}(x, k_\perp,\xi, \Delta_\perp=0) \, .\label{eq1}
\end{eqnarray}
All quantities appear in the above equation will be specified below.
The quark OAM can be reconstructed by integrating over the $x$-dependent OAM distribution: $L_{q}=\int_0^1 dx L_{q}(x,\xi=0)$.  This relation, coupled with Eq.~(\ref{eq1}), thus opens a new avenue to directly access the parton canonical OAM contribution to the nucleon spin through GTMDs. Note that this relation is expected to hold beyond the tree level up to some power corrections~\cite{Ebert:2022cku,Bertone:2022awq,Echevarria:2022ztg}. In recent years, theoretical efforts have primarily centered on investigating the experimental signals of the gluon GTMD $F_{1,4}$~\cite{Ji:2016jgn,Hatta:2016aoc,Bhattacharya:2022vvo,Bhattacharya:2018lgm,Boussarie:2018zwg}.  Conversely, the exclusive double Drell-Yan process, the sole known process providing access to quark GTMDs, mainly offers sensitivity to quark GTMD $F_{1,4}$ in the Efremov-Radyushkin-Brodsky-Lepage (ERBL) region~\cite{Bhattacharya:2017bvs}. This poses a challenge when extrapolating the distribution to the forward limit.

In this paper, we introduce a novel observable to experimentally detect the quark GTMD $F_{1,4}$ in the Dokshitzer–Gribov–Lipatov–Altarelli–Parisi (DGLAP) region, establishing a direct link to quark OAM through Eq.~(\ref{eq1}). Our proposal involves the exclusive $\pi^0$ production process in electron-proton collisions(see Fig.~\ref{illus}): $ep \rightarrow e' p' \pi^0$, with a longitudinally polarized proton target. Our analysis demonstrates that the longitudinal single target-spin asymmetry results in a $\sin 2(\phi_{l_\perp}-\phi_{\Delta_\perp})$ azimuthal angular correlation, where $\phi_{l_\perp}$ and $\phi_{\Delta_\perp}$ denote the azimuthal angles of the transverse momentum of the scattered electron and the recoil proton. This correlation exhibits a direct sensitivity to quark OAM.

The proposed observable stands out as an ideal probe for quark OAM from both theoretical and practical perspectives. Firstly, the background for this process remains clean, free from contamination by final-state soft gluon radiation effects~\cite{Hatta:2021jcd,Hatta:2020bgy,Zhang:2020onw,Tong:2022zwp,Gao:2023ulg,Tong:2023bus}. Additionally, our observable, akin to the unpolarized cross section, constitutes a twist-3 contribution (or equivalently, a sub-leading power correction).  This characteristic enables the maximal enhancement of the asymmetry without being washed out by the unpolarized cross section.
 
\

{\it 2. Probing the quark GTMD $F_{1,4}$ in exclusive $\pi^0$ production}---First, let us define the kinematics of the process under consideration,
 \begin{eqnarray}
 e(l)+p(p,\lambda) \longrightarrow \pi^0(l_\pi)+e(l')+p(p',\lambda') \, .
\end{eqnarray}
The standard kinematic variables are defined as follows: $Q^2=-q^2=-(l-l')^2$, representing the photon's virtuality; and the incoming electron's momentum is parameterized as $l^\mu= (l^+,l^-,l_\perp)=(\tfrac{Q(1-y)}{\sqrt{2}y}, \tfrac{Q}{\sqrt{2}y}, \tfrac{Q\sqrt{1-y}}{y})$. Here, `$+/-$' denotes the light-cone plus/minus components. The $y = p \cdot q / p \cdot l$ represents the usual momentum fraction. The $\gamma^* p$ center-of-mass energy is given by $W^2 = (p + q)^2$. The pion mass in our calculation is neglected ($l_\pi^2 \approx 0$), simplifying the analysis. We work in the symmetric frame where the initial state and the final state proton carry the transverse momenta $p_\perp=-\Delta_\perp/2$ and $p_\perp'=\Delta_\perp/2$, respectively. The skewness variable is given by $\xi = (p^+ - p'^+) / (p^+ + p'^+) = - \Delta^+ / (2 P^+) = x_B / (2 - x_B)$, $x_B = Q^2/2 p \cdot q$, and the momentum transfer squared can be expressed as $t=(p-p')^2=-\tfrac{4\xi^2 M^2+\Delta_\perp^2}{1-\xi^2}$, with $M$ being the proton mass.

In the near forward region, the leading power contribution to the  exclusive transversely polarized virtual photon production  of $\pi^0$ emerges at the twist-3 level.
This suppression of the leading power contribution is a result of the conservation of angular momentum along the direction of the virtual-nucleon beam. Since the unpolarized cross section starts at twist 3, the investigated longitudinal-spin asymmetry is not power-suppressed.
In the region where the momentum transfer $t$ is exceedingly small, the exclusive $\pi^0$ production process becomes susceptible to being dominated by the Primakoff process~\cite{Primakoff:1951iae,Gasparian:2016oyl,Liping:2014wbp,Kaskulov:2011ab,Lepage:1980fj,Khodjamirian:1997tk,Jia:2022oyl}, and the interference between the Primakoff process and the contribution from the gluon GTMD $F_{1,4}$~\cite{Bhattacharya:2023yvo}. In this work, we specifically concentrate on the valence quark region, where $\xi \sim 0.1$, thereby permitting the neglect of contributions from both the Primakoff process and the gluon-initiated process~\cite{Bhattacharya:2023yvo}. 

We will perform the calculation within the framework of collinear higher-twist expansion. This technique, first developed in Refs. ~\cite{Ellis:1982wd,Ellis:1982cd}, was applied to the study of the canonical OAM (see Ref.~\cite{Ji:2016jgn}'s Eq.~(5)), which we closely follow in this work. In this approach, the hard factor $H(k_\perp, \Delta_\perp)$ is expanded in terms of $k_\perp/Q$ and $\Delta_\perp/Q$, where $k_\perp$ denotes the relative transverse momentum carried by the exchanged quarks,
\begin{eqnarray} 
 && \!\!\!\!\!\!\!\!
 H( k_\perp, \Delta_\perp)= H( k_\perp=0, \Delta_\perp=0)+ \label{expansion}
 \\&& \!\!\!\!\! \frac{\partial H( k_\perp,\Delta_\perp\!\!=0)}{\partial k_\perp^\mu}\Big|_{k_\perp\!\!=0}\!\!k_\perp^\mu+ \frac{\partial H( k_\perp\!\!=0,\Delta_\perp)}{\partial \Delta_\perp^\mu}\Big|_{\Delta_\perp \!\!=0}\!\! \Delta_\perp^\mu+... \nonumber
 \label{e:twist3}
\end{eqnarray}
The zeroth-order expansion of $k_\perp$ and $\Delta_\perp$ yields a null result for both the spin-averaged cross section and the longitudinal polarization-dependent cross section. Following this expansion, the subsequent step involves integrating over $k_\perp$. Consequently, the scattering amplitudes are expressed as the convolution of the next-to-leading power of Eq.~(\ref{expansion}) with the GPDs or the first $k_\perp$-moment of certain GTMDs, including the $k_\perp$-moment of the quark GTMD $F_{1,4}$---in other words, the quark OAM distribution.

The leading twist quark GTMDs for nucleons are parameterized as the off-forward  quark-quark correlator~\cite{Meissner:2009ww,Lorce:2013pza},
\begin{equation} \label{e:gtmd_corr}
W_{\lambda,\lambda'}^{ [\Gamma]}  =\!\!
\int \frac{  d^3 z}{2 (2\pi)^3} \, e^{i k \cdot z} \,
\langle p', \lambda' | \, \bar{q}(- \tfrac{z}{2})  \Gamma q(\tfrac{z}{2}) \, | p, \lambda \rangle \Big|_{z^+ = 0} \, ,
\end{equation}
where $\Gamma$ indicates a generic gamma matrix. The Wilson line in Eq.~(\ref{e:gtmd_corr}) is suppressed for brevity.  In the notation of~\cite{Meissner:2009ww,Lorce:2011kd}, they are expressed as follows:
\begin{eqnarray} \label{e:gammap}
W_{\lambda,\lambda'}^{ [\gamma^+]} \! &=&  \!\frac{1}{2M}  \bar{u}(p',\lambda') \bigg[
F_{1,1} + \frac{i  \sigma^{i+} }{P^+}   (k_\perp^iF_{1,2} + \Delta_\perp^i F_{1,3})
\nonumber \\
&&\! + \frac{i \sigma^{ij} k_{\perp}^i \Delta_{\perp}^j}{M^2} \, F_{1,4}  \bigg] u(p,\lambda) \,,
\\[0.2cm]  \label{e:gammap5}
 W_{\lambda,\lambda'}^{ [\gamma^+ \gamma_5]} \!&=&\! \frac{1}{2M}  \bar{u}(p',\lambda') \bigg[
\frac{-i \varepsilon_\perp^{ij} k_{\perp}^i \Delta_{\perp}^j}{M^2}  G_{1,1} + \frac{i  \sigma^{i+}  \gamma_5 k_\perp^i}{P^+}  G_{1,2}
\nonumber \\
& & \! + \frac{i  \sigma^{i+}  \gamma_5 \Delta_\perp^i}{P^+} G_{1,3}
+ i \sigma^{+-} \gamma_5  G_{1,4}  \bigg] u(p,\lambda) \, ,
\end{eqnarray}
where $\varepsilon_\perp^{ij} = \varepsilon^{-+ij}$ with $\varepsilon^{0123} = 1$.
The arguments of the GTMDs depend on $(x, \xi, \vec{k}_\perp, \vec{\Delta}_\perp, \vec{k}_\perp \cdot \vec{\Delta}_\perp)$ but have been omitted in the above formulas for the sake of notation convenience.  In addition to $F_{1,4}$, the quark GTMD $G_{1,1}$ is particularly intriguing. The real part of $G_{1,1}$ encodes information about the quark's spin-orbital angular momentum correlation inside an unpolarized nucleon~\cite{Meissner:2009ww,Lorce:2011kd}. These GTMDs have been explored in various models~\cite{Meissner:2008ay, Meissner:2009ww,Lorce:2011kd,Kanazawa:2014nha,Mukherjee:2014nya,Hagiwara:2016kam,Zhou:2016rnt,Courtoy:2016des,Boer:2023mip,Boer:2021upt,Hatta:2022bxn,Tan:2023kbl,Xu:2022abw,Ojha:2022fls,Ojha:2023etw}, and studied in the small $x$ limit~\cite{Hatta:2022bxn,Agrawal:2023cdf}.

\begin{figure}
    \centering
\includegraphics[scale=0.08]{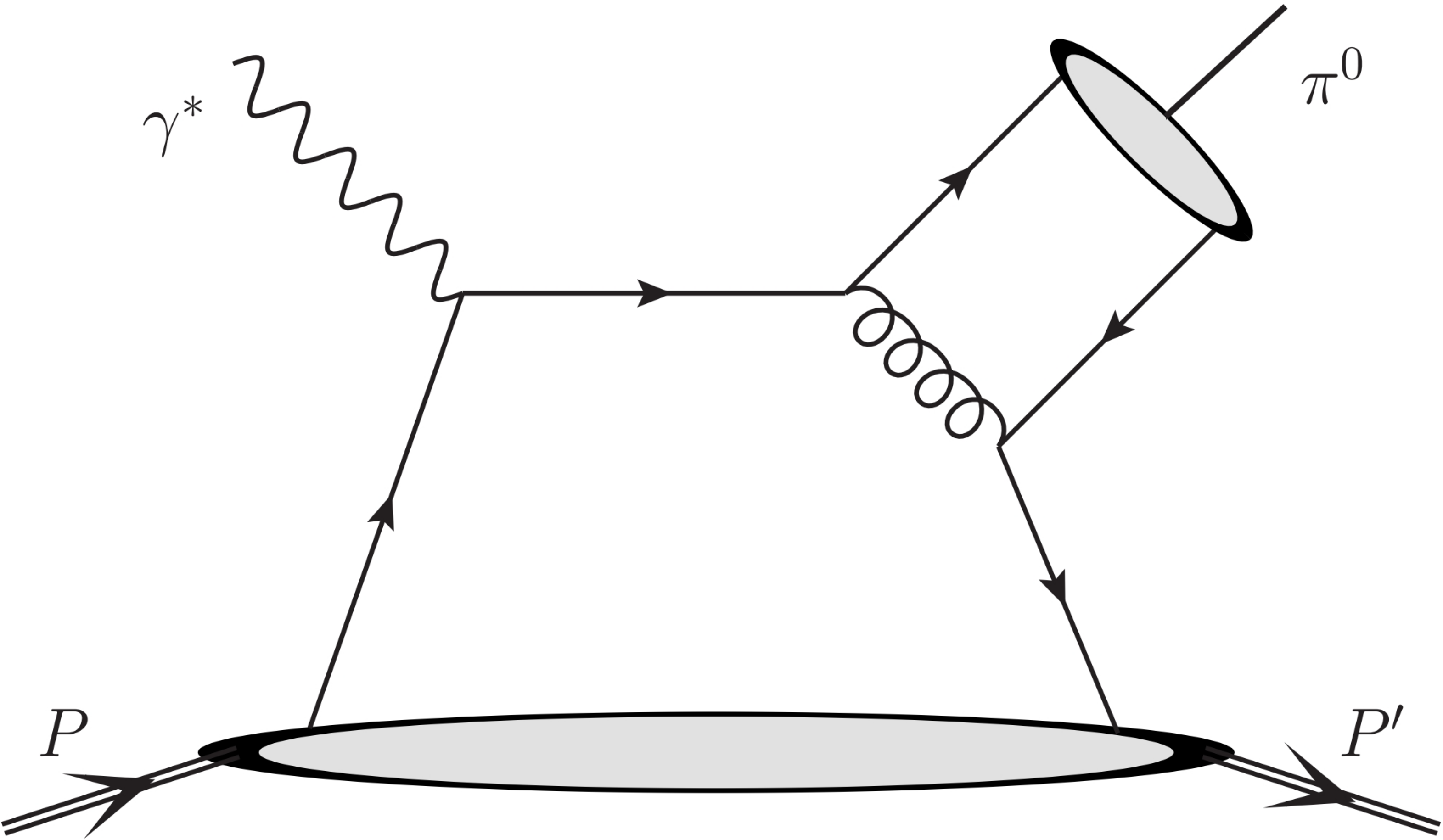}
   \caption{ A diagram contributing to exclusive $\pi^0$ production.}
\label{feynman}
\end{figure}

There are a total of four diagrams contributing to the exclusive $\pi^0$ production amplitude. One of these diagrams is shown in Fig.~\ref{feynman}. Our explicit calculation has confirmed that the contributions from all four diagrams vanish at the leading power. To isolate the twist-3 contribution, we perform an expansion in $k_\perp$ and $\Delta_\perp$. In doing so, it is essential to handle the $k_\perp$ and $\Delta_\perp$ dependencies from the exchanged quark legs with utmost care. To address this, we employ a technique known as the special propagator technique, first introduced in Ref.~\cite{Qiu:1988dn}. The inclusion of the special propagator contribution is crucial to ensure electromagnetic gauge invariance. It is noteworthy that an alternative approach, which also maintains electromagnetic gauge invariance at twist-3 accuracy, has been developed in Refs.~\cite{Anikin:2000em,Radyushkin:2000ap}.  

Depending on the various vector structures, the scattering amplitude can be organized into three terms,
\begin{eqnarray}
{\cal M}_{1}\!\!&=&\!\! \frac{g_s^2e f_\pi }{2\sqrt{2}}   \frac{(N_c^2-1)2\xi}{N_c^2\sqrt{1-\xi^2}}\delta_{\lambda \lambda'}  \frac{  \epsilon_\perp  \times  \Delta_\perp}{Q^2} \left \{ {\cal F}_{1,1} +{\cal G}_{1,1} \right \} ,
\nonumber  \\
{\cal M}_{2}\!\!&=& \!\!\frac{g_s^2e f_\pi }{2\sqrt{2}}\frac{(N_c^2-1)2\xi}{N_c^2\sqrt{1-\xi^2}}\delta_{\lambda,-\lambda'}  \frac{ M \epsilon_\perp  \cdot S_\perp}{ 
 Q^2}\left \{ {\cal F} _{1,2}+{\cal G}_{1,2} \right \} ,
 \nonumber  \\
 {\cal M}_4\!\!&=&\!\! \frac{ig_s^2e f_\pi }{2\sqrt{2}}\frac{(N_c^2-1)2\xi}{N_c^2\sqrt{1-\xi^2}} \lambda \delta_{\lambda \lambda'}  \frac{  \epsilon_\perp  \cdot  \Delta_\perp}{Q^2} \left \{ {\cal  F}_{1,4}+ {\cal G} _{1,4}  \right \} , 
\end{eqnarray}
where $f_\pi=131$ MeV represents the $\pi^0$ decay constant, $\epsilon_\perp$ denotes the virtual photon's transverse polarization vector, and $S_\perp$ is defined as $S_\perp^\mu=(0^+,0^-,-i,\lambda)$. ${\cal F}_{i,j}$ and ${\cal G}_{i,j}$ serve as shorthand notations for complex convolutions involving the GTMDs $F_{i,j}$, $G_{i,j}$, and the $\pi^0$ distribution amplitude (DA) $\phi_\pi(z)$. They are expressed as follows,
\begin{eqnarray} 
{\cal F} _{1,1}\!\!&=& \!\!\int dx   dz \tilde F_{1,1}^{(0)}(x,\xi,\Delta_\perp) x^2 
 \frac{\phi_\pi(z)(1+z^2-z)}{z^2(1-z)^2 } \, , 
\\
{\cal G}_{1,1}\!\!&=& \!\! \int dx   dz    \tilde G_{1,1}^{(1)}(x,\xi,\Delta_\perp)\frac{\phi_\pi(z) ( x^2+2x^2z+\xi^2)}{z^2  }     \, , 
\\
{\cal F} _{1,2}\!\!&=& \!\! \int dx   dz \tilde F_{1,2}^{(1)}(x,\xi,\Delta_\perp) \nonumber \\&&
\times    x \xi (1-\xi^2)
\frac{\phi_\pi(z)(1+z^2-z)}{z^2(1-z)^2 } \, , \\
{\cal G}_{1,2}&=& \int dx   dz  \tilde G_{1,2}^{(1)}(x,\xi,\Delta_\perp)  
\nonumber \\&&
\times  \frac{\phi_\pi(z) ( x^2+2x^2 z+\xi^2  )(1-\xi^2)}{z^2  } \, , 
\\
{\cal F} _{1,4}\!\!&=& \!\!\int dx   dz \tilde F_{1,4}^{(1)}(x,\xi,\Delta_\perp) x \xi  
 \frac{\phi_\pi(z)(1+z^2-z)}{z^2(1-z)^2 } \, ,
\\ 
{\cal G} _{1,4}\!\!&=& \!\! \int dx   dz \tilde G_{1,4}^{(0)}(x,\xi,\Delta_\perp)  \nonumber \\ &&\times  \frac{x(4\xi^2 z +\xi^2-2x^2 z+x^2 ) }{z^2\xi }\phi_\pi(z)    \, ,
\end{eqnarray}
where
\begin{eqnarray} 
\tilde f^{(n)}(x,\xi,\Delta_\perp) \!=\!\int \! \! d^2 k_\perp \! \left( \frac{k_\perp^2}{M^2} \right )^{\!\!n} \! \frac{{  \frac{1}{\sqrt{2}} \left ( \frac{2}{3} f^{u} + \frac{1}{3} f^{d}\right ) }}{ (x+\xi - i \epsilon)^2(x-\xi+i\epsilon)^2} \, \,
\end{eqnarray}
with $n=0,1$, and
$\int dx dz \equiv \int_{-1}^1 dx \int_0^1 dz$. The superscript on the GTMDs $`f$', whose arguments have been suppressed for brevity, indicates the summation of up and down quark contributions.  Here, $z$ represents the longitudinal momentum fraction of $\pi^0$ carried by the quark. The derivation of the above expressions involves the repeated use of the symmetry property: $\int dz \frac{z\phi_\pi(z)}{z^2(1-z)^2} = \int dz \frac{(1-z)\phi_\pi(z)}{z^2(1-z)^2}$.

A few remarks are now in order. First, we obtain the terms ${\cal F}_{1,2}$, ${\cal F}_{1,4}$, ${\cal G}_{1,1}$, and ${\cal G}_{1,2}$ by performing $k_\perp$ expansion, while the $\Delta_\perp$ expansion gives rise to the contributions ${\cal F}_{1,1}$ and ${\cal G}_{1,4}$. Second, the amplitudes ${\cal M}_1$, ${\cal M}_2$, and ${\cal M}_4$ exhibit distinct $\Delta_\perp$-dependent behaviors. Notably, ${\cal M}_2$ persists as $\Delta_\perp$ approaches zero, even when averaging over $S_\perp$ in the unpolarized cross section. This persistence is attributed to the helicity flip mechanism provided by the quark GTMDs $F_{1,2}$ and $G_{1,2}$, akin to what the gluon GTMD $F_{1,2}$ does~\cite{Boussarie:2019vmk}.  The last point to emphasize is that exclusive $\pi^0$ production selects a C-odd exchange. This implies that the hard factors associated with $F_{1,1}$, $G_{1,1}$, and $G_{1,2}$ must be even functions of $x$, while those proportional to $F_{1,2}$, $F_{1,4}$, and $G_{1,4}$ must be odd functions of $x$. This property is explicitly satisfied by our results.   Note that our treatment of the twist-3 contribution to spin independent amplitudes differs from the
approach advocated in Ref.~\cite{Goloskokov:2007nt,Duplancic:2023xrt} which involves a twist-3 pion DA.

Assembling all the pieces, we derive the following spin-averaged and single target longitudinal polarization-dependent cross section:
\begin{eqnarray}
&& \!\!\!\! \!\!\!\! \!\!
\frac{d \sigma_T}{dt dQ^2 dx_B d\phi} \!= \! \frac{(N_c^2-1)^2 \alpha_{em}^2 \alpha_s^2 f_\pi^2  \xi^3\Delta_\perp^2}{ 2 N_c^4 (1-\xi^2) Q^{10}(1+\xi)  } \left [1+\!(1\!-\!y)^2 \right ]
\nonumber \\ &\times& \!\!\! \left \{ \left [ |{\cal F}_{1,1}+{\cal G}_{1,1}|^2 +|{\cal F}_{1,4} +{\cal G}_{1,4}|^2+ \! 2 \frac{M^2}{\Delta_\perp^2} | {\cal F}_{1,2}+{\cal G}_{1,2}|^2 \right ]
 \right .\
\nonumber \\ &&  +\cos (2 \phi) a\left [ - |{\cal F}_{1,1}+{\cal G}_{1,1}|^2 +|{\cal F}_{1,4} +{\cal G}_{1,4}|^2 \right ] 
\nonumber \\ &&  +\lambda \sin (2 \phi) \, 2a \, {\rm Re} \left [\left(  i {\cal F}_{1,4} + i {\cal G}_{1,4} \right ) \left ( {\cal F}^*_{1,1}+{\cal G}^*_{1,1} \right ) \right ]
\Big \} ,
\label{e:main}
\end{eqnarray}
where $\phi=\phi_{l_\perp}-\phi_{\Delta_\perp}$ and $a=\frac{2(1-y)}{1+(1-y)^2}$. Eq.~(\ref{e:main}) stands as the central result of our paper. The real part of the quark GTMD $F_{1,4}$, and consequently, the quark OAM, leaves a distinct signature through an azimuthal angular correlation of $\sin 2\phi$ in the longitudinal single target-spin asymmetry.


\begin{figure}
    \centering
\includegraphics[scale=0.30]{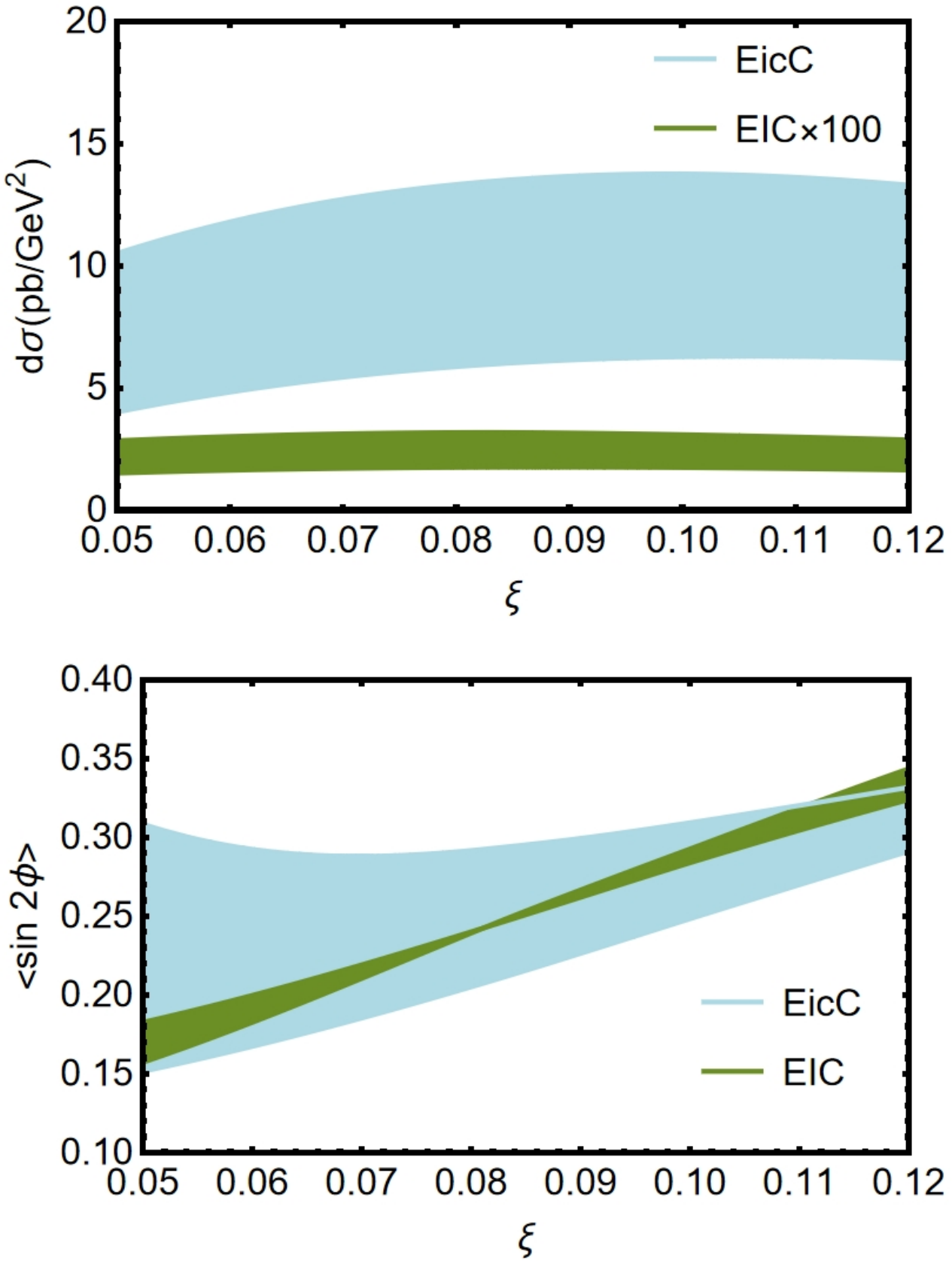}
    \caption{The unpolarized  cross section, as given by Eq.~(\ref{e:main}), is displayed in the top plot for EIC kinematics with $Q^2=10 \, \textrm{GeV}^2$ and $\sqrt{s_{ep}}=100 \, \textrm{GeV}$, as well as for EicC kinematics with $Q^2=3 \, \textrm{GeV}^2$ and $\sqrt{s_{ep}}=16 \, \textrm{GeV}$. The unpolarized cross section for the EIC case is re-scaled by a factor of 100. The bottom plot shows the average value of $\langle \sin(2\phi) \rangle$ given by Eq.~(\ref{e:sin_av_def}). The variable $t$ is integrated over the range [$-0.5\, \textrm{GeV}^2$, $-\frac{4\xi^2 M^2}{1-\xi^2} $].The error bands are obtained by varying the value of $\sqrt{\langle p_\perp^2 \rangle} $ from $150$ MeV to $250$ MeV and the value of $\alpha'$, which determines the $t$-dependence of the various distributions in the double distribution approach (see supplementary material), from 1.2 to 1.4. }
\label{plot}
\end{figure}
{\it 3. Numerical results}---We now present the numerical results for both the unpolarized cross section and the $\sin 2\phi$ asymmetry. It is noteworthy that two of the $k_\perp$-integrated GTMDs, $F_{1,1}$ and $G_{1,4}$, can be linked to the standard unpolarized GPD and the helicity GPD~\cite{Meissner:2009ww}.
In the forward limit, the GTMD $F_{1,2}$ is related to the Sivers function $f^\perp_{1T}$~\cite{Meissner:2009ww,Boussarie:2019vmk,Boer:2015pni,Zhou:2013gsa}, and the GTMD $G_{1,2}$ reduces to the worm-gear function $g_{1T}$~\cite{Meissner:2009ww}. In our first attempt at a numerical study, we choose to neglect contributions from the GTMD $G_{1,1}$, which lacks a GPD or TMD counterpart~\footnote{After submitting this work, a preprint was released providing information about  the small-$x$ behavior of the GTMD $G_{11}$~\cite{Bhattacharya:2024sck,Bhattacharya:2024sno}. We intend to update our numerical analysis in the future to incorporate the contribution of this GTMD.}. Regarding the ${\cal F}_{1,2}$, ${\cal F}_{1,4}$,  and ${\cal G}_{1,4}$ terms, we only consider their pole contributions from their imaginary parts. However, for the term ${\cal F}_{1,1}$, ${\cal G}_{1,2}$,  we include both its imaginary and real parts in the numerical estimation, as they dominate the cross section at high and low $t$ respectively.


Note that the hard part becomes divergent as $z$ approaches 0 or 1. This behavior, known as the endpoint singularity, typically signals factorization breaking. From a phenomenological standpoint, regularization is achievable by considering the transverse momentum dependence of the pion DA~\cite{Goloskokov:2005sd,Goloskokov:2006hr,Sun:2021gmi}. An effective way to introduce transverse momentum dependence is to modify the upper and lower integration limits of $z$ to $\int_{\langle p_\perp^2 \rangle /Q^2}^{1-\langle p_\perp^2 \rangle /Q^2} d z $~\cite{Goloskokov:2007nt}, where $\langle p_\perp^2 \rangle$ is the mean squared transverse momentum of the quark inside the pion. Its central value is chosen to be $\langle p_\perp^2 \rangle = 0.04 \ \textrm{GeV}^2$ in our numerical calculation, based on a fit to the CLAS data (see supplementary material for brief discussion, which includes Refs.~\cite{CLAS:2014jpc,Hand:1963bb,JeffersonLabHallA:2016wye,Mankiewicz:1997uy,Guichon:1998xv,Berthou:2015oaw}). For simplicity, we consider the asymptotic form for the pion's DA, $\phi_\pi (z)=6z(1-z)$. On the other hand, the discontinuity of the derivative of quark GPDs at the endpoints $x=\pm \xi$ (as seen in, for example, Refs.~\cite{Bhattacharya:2018zxi,Bhattacharya:2019cme}), coupled with the double poles at $x=\pm \xi$, may also potentially lead to a divergent component in the cross section. To address this potential issue, we employ a shift of the double pole from $\tfrac{1}{(x-\xi +i\epsilon)^2}$ to $\tfrac{1}{(x-\xi -\langle p_\perp^2 \rangle/Q^2 +i\epsilon)^2}$ (and similarly for the negative $x$ region). A similar shift was introduced in Ref.~\cite{Anikin:2002wg} to handle the aforementioned divergence.  
 More phenomenological input~\cite{Radyushkin:1998es,Radyushkin:2000uy,Goloskokov:2007nt,Hatta:2012cs,Goloskokov:2008ib,Goloskokov:2009ia,Goloskokov:2011rd,Bhattacharya:2021twu,Yang:2024bfz,Echevarria:2014xaa,Bacchetta:2020gko,Echevarria:2020hpy,Bury:2021sue} is detailed in the supplemental material.

We now present numerical predictions for the EIC and EicC kinematics. The $t$-integrated unpolarized cross section is shown as a function of $\xi$ in the top panel of Fig.~\ref{plot}. The asymmetry, quantified by the average value of $\sin (2\phi)$ and depicted as a function of $\xi$ in the bottom plot of Fig.~\ref{plot}, is defined as:
\begin{eqnarray}
\langle \sin(2\phi) \rangle &=&\frac{  \int \frac{d \Delta \sigma}{d {\cal P.S.}}\sin(2\phi) \ d {\cal P.S.} }
{\int \frac{d \sigma}{d {\cal P.S.}}  d {\cal P.S.}} \, ,
\label{e:sin_av_def}
\end{eqnarray}
where $d\Delta \sigma = \sigma (\lambda =1) - \sigma (\lambda =-1) $.
The unpolarized cross section exhibits a notable magnitude at EicC energy, whereas it is relatively small at EIC energy. Note that at EIC, the cross section for low $Q^2$ would be similar to that at EicC. However, EIC's smaller $\xi$ for the same $Q^2$ might offer a greater leverage in constraining quark OAM in the small $x$ region. Additionally, the asymmetries are substantial for both EIC and EicC kinematics. Consequently, our numerical results signify that the azimuthal asymmetry $\sin 2\phi$ in exclusive $\pi^0$ production stands out as a promising avenue for probing the quark OAM distribution.

\ 

{\it 4. Summary}---We propose extracting the quark OAM associated with the Jaffe-Manohar spin sum rule by measuring the azimuthal angular correlation $\sin 2\phi$ in exclusive $\pi^0$ production at EIC and EicC. This observable serves as a clean and sensitive probe of quark OAM for several reasons. Firstly, the azimuthal asymmetry is not a power correction, as both the unpolarized and longitudinal polarization-dependent cross sections contribute at twist-3. Secondly, the produced $\pi^0$ transverse momentum $-\Delta_\perp$ remains unaffected by final state QCD radiations. Detecting $\pi^0$ makes it less challenging to experimentally measure $\Delta_\perp$, in contrast to the diffractive di-jet production case where reconstructing $\Delta_\perp$ from the total transverse momentum of the di-jet system is impossible due to the contamination of final-state soft gluon radiations. Most importantly, this process enables the {\it direct} access to the quark GTMD $F_{1,4}$ in the DGLAP region for the first time.  In addition to unveiling access to quark OAM, our work highlights another significant finding that the quark component of $F_{1,2}$ and $G_{1,2}$, or equivalently the Sivers function and the worm-gear function respectively, contribute to the unpolarized cross-section of this process. This result is particularly noteworthy since conventionally, the Sivers function and the worm-gear function are  understood to be probed only through transversely polarized targets.

We computed the differential cross section within the collinear higher-twist expansion framework. Despite the substantial uncertainties associated with the model inputs, our numerical results reveal a sizable azimuthal asymmetry, which critically relies on the quark OAM distribution. In the kinematic range accessible to the EIC and EicC, our observable can be thoroughly investigated, paving the way for the first experimental extraction of the canonical quark OAM distribution in the future.

\

{\it Acknowledgements:} The authors would like to express their gratitude to C. Cocuzza for providing the LHAPDF tables of JAM22 PDFs as referenced in~\cite{Cocuzza:2022jye}, and to Ya-ping Xie for sharing code with us from Refs.~\cite{Goloskokov:2022rtb,Goloskokov:2022mdn,Xie:2023mch}. We  also thank  Y. Hatta and F. Yuan for their insightful discussions. This work has been supported by  the National Natural Science Foundation of China under Grant No.~12175118 and  under Contract No.~PHY-1516088(J.~Z.). S.~B. has been supported by the U.S. Department of Energy under Contract No. DE-SC0012704, and also by  Laboratory Directed Research and Development (LDRD) funds from Brookhaven Science Associates. 

\newpage

\begin{widetext}

\setcounter{equation}{0}
\setcounter{figure}{0}
\setcounter{table}{0}

\parskip=6pt

{\bf \centerline{SUPPLEMENTARY MATERIAL}}
In the supplementary material of our paper, we present more details of  numerical estimations and compare our theoretical calculations with the CLAS measurement of the unpolarized exclusive $\pi^0$ production cross section~\cite{CLAS:2014jpc}.

\subsection{ Phenomenological inputs for numerical estimations}
To provide a model input for the $x$-dependent quark OAM distribution, specifically the $k_\perp$-moment of $F_{1,4}$, we employ the Wandzura-Wilczek (WW) approximation~\cite{Hatta:2012cs}, 
\begin{eqnarray}
L_q(x)\approx x \! \int_x^1 \! \frac{dx'}{x'} \big (H_q(x')+E_q(x')\big )-x \! \int_x^1 \! \frac{dx'}{x'^2}{\tilde{H}_q(x')} \, ,
\label{e:oam_ww}
\end{eqnarray}
where $H_q(x')=q(x')$ and $\tilde{H}_q(x')=\Delta q(x')$ denote the usual unpolarized quark PDF and quark helicity PDF, respectively. This approximation, neglecting genuine twist-3 terms, is often employed for convenience in the absence of reliable experimental estimations. We use the JAM (valence) quark PDFs $q(x)$ and $\Delta q(x)$ as inputs in Eq.~(\ref{e:oam_ww}) from Ref.~\cite{Cocuzza:2022jye}. As for GPD $E_q(x')$, we use the parameterization from \cite{Goloskokov:2008ib} and \cite{Goloskokov:2009ia}.
It is a common practice to reconstruct the $\xi$-dependence for $x L_q(x,\xi)$ from its PDF counterpart $x L_q(x)$ using the double distribution method~\cite{Radyushkin:1998es,Radyushkin:2000uy,Goloskokov:2007nt}. Specifically, we use:
\begin{eqnarray}
F_q(x, \xi)=\int_{-1}^1 d \beta \int_{-1+|\beta|}^{1-|\beta|} d \alpha \delta(\beta+\xi \alpha-x) f_q(\alpha,\beta) 
\end{eqnarray}
where 
\begin{eqnarray}
f_q( \alpha,\beta) &= &  \frac{3}{4} \frac{\left[(1-|\beta|)^2-\alpha^2\right]}{(1-|\beta|)^{3}}  \beta 
L_q (\beta) \, .
\end{eqnarray}
As for its $t$-dependence, we adopt a Gaussian form factor, represented by $e^{t/\Lambda}$ with $\Lambda = 0.5 \, \text{GeV}^2$.

The Compton form factors in Eqs.~(10)-(11) involve the $k_\perp$ moments of GTMDs $F_{1,2}$ and $G_{1,2}$.  These moments can be determined in the forward limit by connecting them to the corresponding TMDs.
Specifically, in this limit, GTMD  $F_{1,2}$ is related to the Sivers function $f^\perp_{1T}$, while GTMD $G_{1,2}$ reduces to the worm-gear function $g_{1T}$.  For numerical estimations,
we rely on the parametrizations  for the quark Sivers function  from Ref.~\cite{Echevarria:2014xaa} and the parametrizations for the worm-gear function  from Ref.~\cite{Bhattacharya:2021twu}.
For $G_{1,2}$ contribution, we utilize the relation,
\begin{eqnarray}
&&\int d^2 k_\perp \frac{k_\perp^2}{2M^2} {\text Re} [G_{1,2}(x,\xi=0,\Delta_\perp=0,k_\perp)]\nonumber \\&&=\int d^2 k_\perp \frac{k_\perp^2}{2M^2} g_{1T}(x,k_\perp) =g_{1T}^{(1)}(x)
\end{eqnarray}
where $g_{1T}^{(1)}(x)$ is parameterized as~\cite{Bhattacharya:2021twu},
\begin{eqnarray}
g_{1T}^{(1)}(x)=\frac{N_q}{\tilde N_q}x^{\alpha_q }(1-x)^{\beta_q } q(x)
\end{eqnarray}
with $\alpha_u=\alpha_d=1.9$, $\beta_u=\beta_d=1$ and $N_u=0.033, N_d=-0.002$. $\tilde N$ is determined through $\tilde N=\int_0^1 dx x^{\alpha+1}(1-x)^\beta q(x)$. We notice that other fitting for $g_{1T}^{(1)}(x)$ exist too~\cite{Yang:2024bfz}.  Meanwhile, the $k_\perp$ moment of $F_{1,2}$ can be related to the Qiu-Sterman function,
\begin{eqnarray}
&&\int d^2 k_\perp \frac{k_\perp^2}{M} {\text Im} [F_{1,2}(x,\xi=0,\Delta_\perp=0,k_\perp)]\nonumber \\&&=-\int d^2 k_\perp \frac{k_\perp^2}{M} f_{1T}^\perp (x,k_\perp) =T_F(x,x)
\end{eqnarray}
where the Qiu-Sterman function is parametrized as~\cite{Echevarria:2014xaa},
\begin{eqnarray}
T_F(x,x)=N_q \frac{(\alpha_q+\beta_q)^{(\alpha_q+\beta_q)}}{\alpha_q^{\alpha_q} \beta_q^{\beta_q} } x^{\alpha_q}(1-x)^{\beta_q} q(x)
\end{eqnarray}
with $\alpha_u=1.051, \alpha_d=1.552$, $\beta_u=\beta_d=4.857$,  and $N_u=1.06, N_d=-0.163$.
See also Refs. \cite{Bacchetta:2020gko}, \cite{Echevarria:2020hpy}, and \cite{Bury:2021sue} for the state-of-the-art extractions of the Sivers functions. Once the $x$-dependence of the $k_\perp$ moments of $F_{1,2}$ and $G_{1,2}$ is reconstructed as explained above, we reconstruct their $(\xi,t)$-dependence in accordance with the double distribution method. Specifically, we use the parameterization:
\begin{eqnarray}
F_q(x, \xi, t)=\int_{-1}^1 d \beta \int_{-1+|\beta|}^{1-|\beta|} d \alpha \delta(\beta+\xi \alpha-x) f_q(\beta, \alpha, t) 
\end{eqnarray}
where 
\begin{eqnarray}
f_q( \alpha,\beta, t) &= &  \frac{3}{4} |\beta|^{-\alpha't}\frac{\left[(1-|\beta|)^2-\alpha^2\right]}{(1-|\beta|)^{3}} 
\begin{cases}
g^{(1)}_{1T}(|\beta|) \nonumber \\
T_{F}(|\beta|, |\beta|)
\end{cases}
\end{eqnarray}
with $\alpha'=1.3$.


Finally, we reconstruct the $\xi$ and $t$-dependence of the GPDs $H(x,\xi, t)$ and $\tilde{H}(x,\xi,t)$ entering Eqs.~(8) and~(13) from their PDF counterparts using the double distribution method.

\subsection{Comparison with the CLAS data}
We now compare our theoretical calculations with the CLAS measurement of the unpolarized exclusive $\pi^0$ production cross section~\cite{CLAS:2014jpc}. The CLAS measurements primarily cover the large skewness region, making it challenging to extrapolate the functional form of GPDs to the forward limit. Nevertheless, the comparison with CLAS data serves as a valuable test for our theoretical calculations and allows us to fine-tune relevant parameters.

The CLAS data is provided for  the reduced or ``virtual photon" cross sections. By adopting the hand convention~\cite{Hand:1963bb} for the virtual photon flux definition, the reduced cross section $d \sigma_T/dt$ can be extracted directly from  Eq.~(15) in the main text,
\begin{eqnarray}
   &&\!\!\!\!\!\!\!\!\!\!\! \frac{d\sigma_T}{dt}
    =\frac{2\pi^2\alpha_s^2\alpha_{em} f_\pi^2(N_c^2-1)^2\xi^3\Delta_\perp^2 x_B}{N_c^4(1-\xi^2)(1+\xi)(1-x_B) Q^8  } \\
    &&\!\!\!\!\!\!\!\!\! \times \! \left [ |{\cal F}_{1,1}+{\cal G}_{1,1}|^2 +|{\cal F}_{1,4} +{\cal G}_{1,4}|^2+ \! 2 \frac{M^2}{\Delta_\perp^2} | {\cal F}_{1,2}+{\cal G}_{1,2}|^2 \right ] \, .\nonumber 
\end{eqnarray}
After accounting for the mass correction which is necessary at the CLAS energy, the above formula is modified as,
\begin{eqnarray}
  &&\!\!\!\!\!\!\!\!\!\!\! \frac{d\sigma_T}{dt}
    =\frac{2\pi^2\alpha_s^2\alpha_{em} f_\pi^2(N_c^2-1)^2\xi^3\Delta_\perp^2}{N_c^4(1-\xi^2)(1-\xi) Q^6\sqrt{\Lambda(W^2,-Q^2,m^2)}  } \\
    &&\!\!\!\!\!\!\!\!\! \times \! \left [ |{\cal F}_{1,1}+{\cal G}_{1,1}|^2 +|{\cal F}_{1,4} +{\cal G}_{1,4}|^2+ \! 2 \frac{M^2}{\Delta_\perp^2} | {\cal F}_{1,2}+{\cal G}_{1,2}|^2 \right ] \nonumber 
\end{eqnarray}
where $\Lambda(x,y,z)$ is defined as $\Lambda(x,y,z)=(x^2+y^2+z^2)-2xy-2xz-2yz$ and $W^2$ is the center of mass energy of $\gamma^*p$ system.
The  reduced cross sections measured by the CLAS Collaboration is given by the combination $\frac{d \sigma_T}{dt}+a \frac{d \sigma_L}{dt} $ where $\sigma_L$ denotes the cross section of the $\pi^0$ production initiated by the longitudinally polarized virtual photons.  The parameter $a$, defined in the main text, represents the ratio of fluxes between longitudinally and transversely polarized virtual photons. It is important to emphasize that the contributions from $\sigma_L$ and $\sigma_T$ can be disentangled by examining the combined cross section at different values of $y$~\cite{JeffersonLabHallA:2016wye}.
\begin{figure}
    \centering
\includegraphics[scale=0.4]{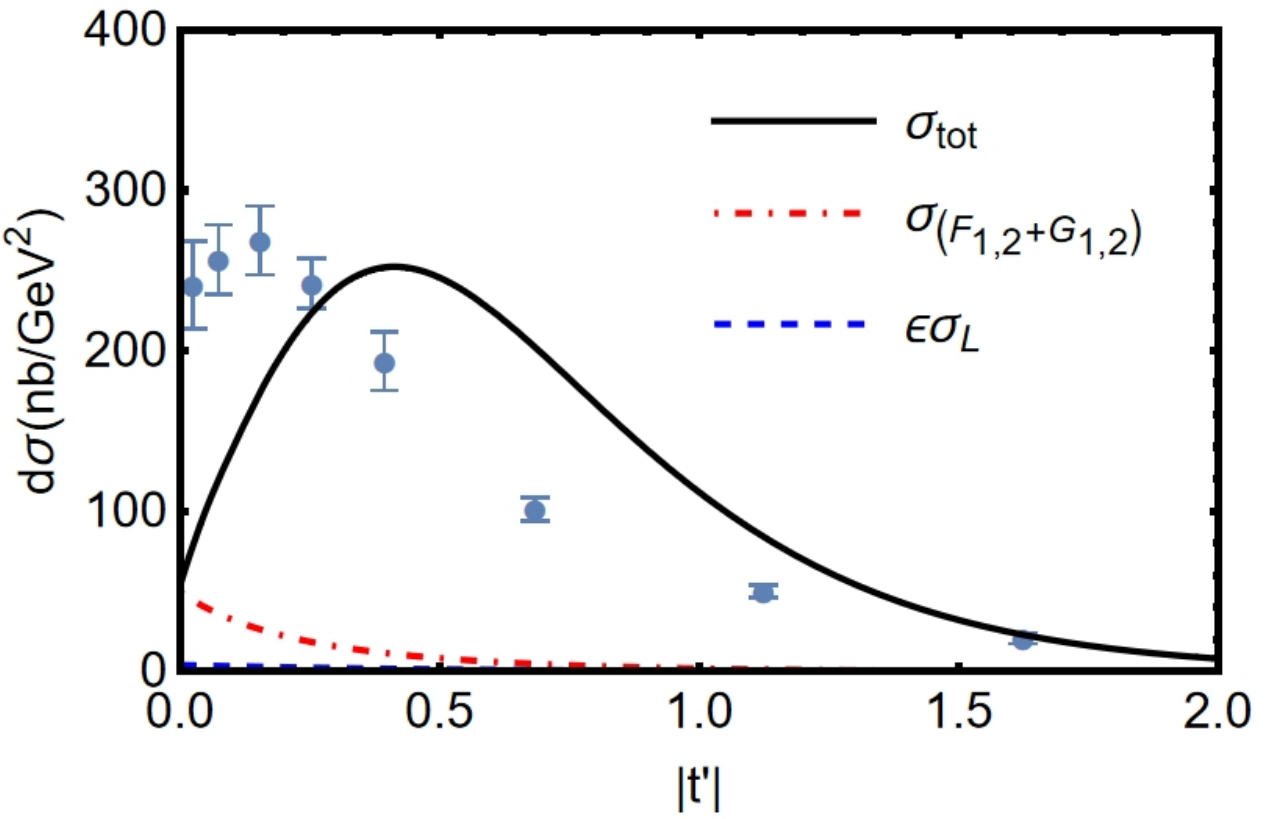}
   \caption{ The unpolarized cross section is plotted as the function of $|t'|$ with $Q^2=2.21 \text{GeV}^2$ and $x_B=0.28$.  The CLAS data are displayed for comparison.}
\label{crosssection}
\end{figure}
 In contrast to exclusive meson production initiated by transversely polarized virtual photons, the factorization for longitudinally polarized photon production has been rigorously proven ~\cite{Collins:1996fb}.  While $\sigma_T$ receives the leading contribution from the twist-3 level, the calculation of  $\sigma_L$ is well formulated  within the leading twist collinear factorization framework~\cite{Mankiewicz:1997uy,Guichon:1998xv}.  However, due to the cancellation between contributions from $u$-quark and $d$-quark  GPDs~\cite{Goloskokov:2009ia,Goloskokov:2011rd}, the yield of 
 the longitudinal photon production of $\pi^0$ was shown to be negligibly small~\cite{Goloskokov:2022rtb,Goloskokov:2022mdn,Xie:2023mch}.

To evaluate $\sigma_L$, one can utilize existing code such as the PARtonic Tomography Of Nucleon Software (PARTONS)~\cite{Berthou:2015oaw}, or the package developed in the Refs.~\cite{Goloskokov:2022rtb,Goloskokov:2022mdn,Xie:2023mch}, which we employ in this study. We plot $\frac{d \sigma_T}{dt}$ and $\frac{d \sigma_L}{dt} $ as  functions of $|t'|=|t+\frac{4\xi^2 M^2}{1-\xi^2} |$ in  Fig.~\ref{crosssection} separately. Additionally, we compare the combined cross section $\frac{d \sigma_T}{dt}+a \frac{d \sigma_L}{dt} $ with the CLAS measurement, as shown in Fig.~\ref{crosssection}. Our calculation is in reasonable agreement with the experimental data.  We observe that $G_{1,2}$ term dominates the cross section as $\Delta_\perp$ approaches zero. In an extended version of this paper, we will conduct a more thorough analysis of the JLab measurements~\cite{CLAS:2014jpc,JeffersonLabHallA:2020dhq} by taking into account another type twist-3 contributions as well~\cite{Goloskokov:2007nt,Duplancic:2023xrt}.

\vspace{0.4cm}

\end{widetext}

\bibliography{ref.bib}

\end{document}